\pdfoutput=1
\documentclass[sigconf,nonacm=true]{acmart}

\AtBeginDocument{%
  }

\setcopyright{none}
\usepackage{tabularx}
\usepackage{enumitem}
\setlist{nosep,leftmargin=1.4em}
\usepackage[most]{tcolorbox}
\usepackage{tikz}
\usetikzlibrary{positioning,arrows.meta,calc,decorations.pathreplacing}

\DeclareRobustCommand{\code}[1]{\texttt{\small #1}}

\newtcolorbox{cidrbox}[1]{
  colback=gray!8,
  colframe=gray!50,
  arc=2mm,
  boxrule=0.4pt,
  left=8pt, right=8pt, top=6pt, bottom=6pt,
  before skip=10pt, after skip=10pt,
  fonttitle=\bfseries,
  title={#1},
  fontupper=\small
}

\newcommand{\principle}[1]{%
  \begin{quote}\itshape\small #1\end{quote}%
}


\begin{document}

\title{Eigenius: A Typed Knowledge-Graph DBMS with Epistemic Stratification and Institution-Mediated Reasoning}

\author{Hans-Martin Will}
\affiliation{%
  \institution{The Eigenius Project}
  \country{United States}}

\author{A. L. Brown Jr.}
\affiliation{%
  \institution{Docimion}
  \country{United States}}

\author{Matthew Fuchs}
\affiliation{%
  \institution{Docimion}
  \country{United States}}

\renewcommand{\shortauthors}{Will et al.}

\begin{abstract}
\noindent
As ``AI Scientists'' emerge to drive research via the Model 
Context Protocol (MCP), systems relying on ephemeral scripts will fail. 
The sheer scale of stateful, interconnected evidence requires a 
machine-walkable warranty grounded in a purpose-built database architecture. 
Eigenius is an open-source, typed knowledge-graph DBMS built on a single premise: 
answering the audit question (``what do you know, and what is your warranty?'') 
requires a unified kernel. By tightly coupling the type system, storage engine, 
and integration protocol, Eigenius turns data provenance into a structural 
invariant rather than a property reconstructed across subsystem boundaries. 
The kernel rests on three pillars whose metatheory is mechanized in Lean~4:
a dependent type theory (\emph{EigenTT}) separating confluent conversion from
ontology subsumption and keeping proof irrelevance axiomatic so distinct warrants
remain inspectable; proof-calculus institutions acting as strongly typed
integration boundaries; and a content-addressed immutable storage layer.
Epistemic status (\emph{declared}/\emph{observed}/\emph{derived}/\emph{verified})
forms a bounded lattice over a justification logic enforced at commit.
Cross-system comorphisms carry explicit axiom policies (\emph{Adopt},
\emph{Discharge}, \emph{Relativize}), transport witnesses at commit, and
materialize outputs into the graph, while EigenQL provides a Grothendieck query
view across all hosted logics and shared on-chain IRs collapse multi-system
translations to identity. Pairing empirical justification logic with an
in-process Lean~4 term checker, an end-to-end recomputation of a published
\emph{Nature} study from pinned data discharges all 52 derived conclusions and
surfaces four machine-checked discrepancies in the original study.
\end{abstract}

\keywords{Knowledge graphs; epistemic provenance; institution
  theory; Grothendieck construction; formal verification; Lean 4;
  dependent type theory; typed merge; content-addressed storage; MCP}

\maketitle

\section{Introduction}
\label{sec:intro}

Scientific data consumption is shifting. As end-to-end ``AI Scientists'' 
(e.g., OpenAI Rosalind, Claude Science) emerge, autonomous agents are driving 
research loops, firing thousands of inferences via the Model Context 
Protocol (MCP)~\cite{mcp}. However, current data infrastructure is unready. 
Today, scientific arguments are captured as ephemeral webs of linked scripts 
and narrative prose. Manual auditing across these fragile pipelines was once 
inconvenient but possible for humans. For an autonomous agent, a 
cryptographically secure, fully walkable proof chain is a load-bearing pillar. 
Formalizing this requires managing millions of stateful, interlinked concepts---from 
gigabyte-scale matrices to granular logical warrants---demanding a purpose-built 
database architecture.

The core ambition of a machine-auditable scientific foundation is not new. 
In 2008, Pharos~\cite{brown2008pharos} articulated a clear vision: 
enforcing the scientific method by making inference and workflow 
machine-auditable. Pharos shipped a credible prototype, eventually productized 
as \emph{Microsoft Amalga Life Sciences 2009}~\cite{amalga}. However, 
that vision outpaced its infrastructure. Without a trusted kernel 
acting as a switchboard, the system relied on fragile peer-to-peer bridges---stitching 
together SQL Server, DataNet via TSVs and .NET stored procedures, and the HOL 
theorem prover via F\#. Lacking a formal protocol like institutions to integrate 
these distinct forms of reasoning, there was no central place for a unifying 
type theory to live. While the mathematical theories required to solve this 
long predate Pharos, their necessity as load-bearing database infrastructure 
was not yet appreciated. Eighteen years later, synthesizing them into a unified 
engine has become critical.

We introduce \textbf{Eigenius}, an open-source\footnote{Prototype and mechanized metatheory: \url{https://github.com/eigenius/eigenius} and \url{https://github.com/eigenius/theory}} typed knowledge-graph DBMS 
that revisits this architecture from the ground up, with its kernel metatheory
mechanized in Lean~4~\cite{will2026theory}. A purpose-built kernel---wholly
owning the type system, storage layer, and integration protocol---provides
structural guarantees that a loosely coupled stack cannot match: integrating
dependent type theory, institutional proof calculi, and content-addressed
storage turns the audit story into a structural invariant of the data model.

\paragraph{Contributions.}

\begin{description}[style=unboxed,leftmargin=0em]
\item[C1.] A self-describing typed-graph data model whose dependent type theory
  (\emph{EigenTT}, mechanized in Lean~4~\cite{will2026theory}) compiles ontology
  classes and constraints into record, refinement, and singleton types while
  keeping proof irrelevance axiomatic so distinct warrants survive conversion for
  audit. Epistemic stratification (\emph{declared}/\emph{observed}/\emph{derived}/\emph{verified})
  forms a bounded lattice over a justification logic, paired with an in-process
  Lean~4 term checker without IPC overhead.
\item[C2.] A three-level institutional protocol~\cite{goguen1992,meseguer1989,will2026theory}
  generalizing the stored procedure to typed knowledge graphs: Meseguer proof
  calculi on the commit path transport witnesses across comorphisms under
  explicit axiom policies (\emph{Adopt}, \emph{Discharge}, \emph{Relativize})
  and materialize outputs into the chain, while EigenQL provides a common
  Grothendieck query view~\cite{diaconescu2025} across all registered institutions.
\item[C3.] A solution to the $O(N^2)$ polystore adapter bottleneck: lifting a
  shared, strongly typed intermediate representation (IR) into the graph's schema
  collapses comorphisms between systems sharing that IR to pure identity.
\item[C4.] An end-to-end recomputation and re-encoding of a published \emph{Nature}
  study from ephemeral scripts to a materialized evidence graph, where 52 of 52
  derived conclusions hold from pinned data, surfacing four discrepancies.
\end{description}

\section{Pharos and the Agent Inversion}
\label{sec:pharos}

To understand Eigenius's design, it is useful to examine the architectural 
ceiling hit by Pharos~\cite{brown2008pharos}. Pharos attempted to build a 
machine-auditable database by federating a graph store (DataNet) with 
probabilistic reasoning (Infer.NET) and formal proofs (MeTaL). This 
federated approach exposed a fundamental limit: each subsystem had its own 
typing and durability story. Schema lived in a different ontological category 
than instance data; bridges were opaque custom code; and standard storage 
recursion could not express institution-dispatched queries that both read 
and write the chain mid-evaluation. 

Eigenius replaces this federated stack with a unified mathematical kernel 
(Table~\ref{tab:pharos-eigenius}).

\begin{table*}[t]
\centering
\caption{Pharos (2008) vs.\ Eigenius (2026), by architectural concern.}
\label{tab:pharos-eigenius}
\small
\begin{tabularx}{\textwidth}{@{}l X X@{}}
\toprule
\textbf{Concern} & \textbf{Pharos (2008)} & \textbf{Eigenius (2026)} \\
\midrule
Schema realization
  & Compiled to SQL Server tables; instances stored as rows
  & Schema \emph{is} data: classes compile to record/refinement types with width subtyping; resource and type form a $\Sigma$-pair $(T,t)$ \\
Storage substrate
  & SQL Server; query planning inherited
  & RocksDB KV store; content-addressed immutable layers; four typed merges \\
Recursion / query
  & SQL Server TVFs driven by SPARQL compilation
  & Kernel-owned seminaive-fixpoint evaluator over \code{DEFINE}d rules \\
Shared payload language
  & None; bespoke per-pair translation
  & Chain-mirrored EigenTT fragment; comorphisms collapse to identity \\
Reasoning integration
  & Bespoke bridges (MeTaL$\leftrightarrow$DataNet); proof translation is infrastructure
  & Proof-calculus comorphisms~\cite{meseguer1989,will2026theory} checked at commit (\emph{Adopt}, \emph{Discharge}, \emph{Relativize}); Grothendieck query view in EigenQL \\
Evidence model
  & A conclusion's support lives inside MeTaL or Infer.NET as engine state
  & Statistics institution recomputes results; reasoning institution checks justification terms~\cite{artemov-jl} whose leaf support induces a four-grade lattice \\
Primary user
  & Human researcher; audit by inspection
  & Designed for AI agents over MCP; the chain is the audit substrate \\
User surface
  & SPARQL window, graph viewer, VS-style workbench
  & MCP tools for agents; React notebook for humans; TypeScript SDK \\
\bottomrule
\end{tabularx}
\end{table*}

\paragraph{The agent inversion.} The last row of 
Table~\ref{tab:pharos-eigenius} highlights the shift making this unified 
architecture mandatory: the transition from human researchers to autonomous 
AI agents. A coding agent converges on a correct program through a tight 
loop with a compiler and linter, turning errors into localized, machine-checkable 
signals. An agent reasoning in prose has no such loop. Supplying one for 
\emph{scientific thinking} motivates what follows: capturing knowledge as typed 
resources makes the structural type checker the agent's compiler, while 
commit-time epistemic gates act as its test suite. Closing that loop with an 
agent reasoning natively through the substrate is future work; this paper 
reports the database kernel required to run it.

\section{Data Model and Epistemic Stratification}
\label{sec:datamodel}

Everything in Eigenius is a \emph{Resource} (IRI identity, class-membership 
claims, typed properties), extending Atomic Data~\cite{atomicdata} with a 
dependent type system and institution machinery. The ontology is self-describing; 
schema and data are unified, so declaring a class simply means committing a resource. 
The unit of commit is a \emph{layer}---an immutable resource set with a parent 
pointer. Layer identity is content-addressed (SHA-256 of canonical CBOR~\cite{rfc8949}), 
forming a Merkle commitment to graph history where class resolution walks the chain.

\paragraph{Why a database type theory differs from a proof assistant's.}
The kernel's native calculus, \emph{EigenTT}~\cite{will2026theory}, is a dependent
type theory with a cumulative universe hierarchy ($\mathrm{Prop} = \mathrm{Sort}\,0 <
\mathrm{Set} = \mathrm{Sort}\,1 < \mathrm{Type}_k$ with impredicative $\mathrm{Prop}$),
inductive families, and intensional identity types. Serving as a persistent
knowledge-graph kernel rather than an interactive proof assistant requires two
departures from calculi such as Lean's~\cite{lean4}:
\begin{enumerate}[leftmargin=1.4em]
\item \emph{Conversion separated from ontology subsumption.} Definitional equality
  ($\Gamma \vdash A \equiv B$) is joinability under an orthogonal rewriting system---confluent,
  injective on type formers, and decided by untyped normalization by evaluation
  (NbE)~\cite{nbe}. Subtyping ($\Gamma \vdash A \leq B$) is a separate judgment
  combining universe cumulativity, codomain-covariant $\Pi$-types, record width
  subtyping, and ontology class inclusions ($\sqsubseteq^*$). Closed-world
  classes compile to record types $\mathrm{Rec}(\vec{\ell} : \vec{A})$, while
  constraints use refinements $\mathrm{Refine}(R, \vec{S})$, intersections
  $A \sqcap B$, and singletons $\mathrm{Sing}_A(a)$.
\item \emph{Axiomatic rather than definitional proof irrelevance.} Proof assistants
  make $\mathrm{Prop}$ \emph{definitionally} proof-irrelevant so any two proofs of
  $P$ convert to the same normal form. In an epistemic database whose job is to
  answer \emph{``with what warranty?''}, collapsing two warrants for $P$ under
  conversion would erase provenance at the kernel level and reduce the proof
  calculus to boolean entailment~\cite{will2026theory}. EigenTT keeps conversion
  on $\mathrm{Prop}$ intensional and supplies proof irrelevance only as an
  explicit axiom ($\mathrm{irrel}$), so distinct proof and justification terms
  remain distinct, inspectable chain resources.
\end{enumerate}

Branches and merges make the layer history a lattice whose joins are typed 
merges (Witness, Rename, SchemaQuotient, Restructure), acting as chain-level 
functorial data migrations~\cite{spivak2012} (theory morphisms in $\mathbf{Th}$).

\paragraph{Epistemic stratification as a justification logic.} A prover discharges a
mathematical claim because its leaves are axioms; a scientific
claim's leaves are observations and conventions --- an instrument
reading, a chosen $\alpha$ --- that no prover can discharge. So the
kernel checks the composition while the grade records what stands
behind each leaf: a justification logic $\mathcal{J}$~\cite{artemov-jl,will2026theory}
whose warrants are deliberately not all factive. Justification terms $j$ are
built from atomic leaves---declarations $\mathsf{decl}(c, P)$, observations
$\mathsf{obs}(e, P)$, and verified proofs $\mathsf{ver}(w, P)$---closed under
modus ponens ($j_1 \cdot j_2$), evidence combination ($j_1 + j_2$), and
instantiation ($j \Downarrow t$). Compositional evidence $\mathcal{E}(j)$ and
support $\mathrm{supp}(j, P)$ verify that $j$ discharges $P$. Only
$\mathsf{ver}(w, P)$ is factive ($\vdash P$); $\mathsf{decl}$ and $\mathsf{obs}$
record non-factive assumptions without asserting truth in the empty context,
preventing contradictory observations from trivializing the store. Aggregating
leaf support induces a four-grade bounded lattice enforced at commit via
structural base classes: \emph{declared} (\code{Declared\-Resource}, authority
checked for well-formedness), \emph{observed} (\code{Observed\-Resource},
external provenance), \emph{derived} (\code{Derived\-Resource}, produced by a
typed computation over warranted premises), and \emph{verified}
(\code{Verified\-Resource}, carrying a formally re-checked proof term).

The protocol that moves a resource between categories is the same
protocol the system uses to plug in reasoning. Institutions declare 
\emph{commit-time triggers}: the kernel fires matching queries on every commit. 
The handler returns a typed \code{Verdict} (\code{Holds} / \code{Fails}); 
promotion to \emph{verified} means ``the formal verifying institution returned 
\code{Holds}.'' The verdict itself is chain-resident.

\paragraph{Scope.} Eigenius's audit chain is \emph{re-walkable}:
an auditor replays the recorded warrants from the graph alone.
Making it \emph{re-runnable} --- governing the processes that
produce warrants, not merely recording them --- is the companion
system \emph{KlinikOS}~\cite{brown2026pct}, described separately.

\paragraph{Verification as a dependent pair.}
DataNet kept a type and its inhabitants in two referentially-linked
tables. In dependent type theory, that referential integrity \emph{is} a
$\Sigma$-type: a pair $(T,t)$ with $t : T$, which Eigenius makes
first-class as a single resource. Because EigenTT represents 
logical propositions natively, the kernel validates a claim's proposition 
$P : \mathrm{Sort}\,0$ (\code{Prop}) strictly upon standard ingestion. 
However, elevating a claim to \emph{verified} status requires committing 
it as a fully inhabited dependent pair $(P, t)$. While empirical reasoners 
supply structural terms to produce \emph{derived} resources (\S\ref{sec:impl}), 
a formal institution uses Lean~4 to supply a verbatim proof payload for $t$. 
Only for claims seeking \emph{verified} status does the kernel invoke its 
in-process verifier to statically re-check this opaque payload. The payload 
meets the kernel through an export comorphism $\iota$ relabeling the 
proposition, ensuring $P = \iota\,T$. At commit, the kernel checks the native 
proposition it owns, leaving the heavy proof obligation to the institution.

\principle{The regulator's question, ``how do you know X is
  safe?'', is a typed query against the chain.}

\section{Institutions, Comorphisms, and EigenQL}
\label{sec:institutions}

To integrate external reasoners (ODE solvers, theorem provers) into a DBMS, 
the standard tool is the stored procedure. Because stored procedures assume a 
flat relational world, Eigenius introduces \emph{institutions}~\cite{goguen1992} 
as a strongly typed generalization for knowledge graphs.

\paragraph{Three institutional levels and witness transport.}
Classical Goguen--Burstall institutions~\cite{goguen1992} and
$\pi$-institutions~\cite{fiadeiro-sernadas1988} model satisfaction ($\models$)
or consequence ($\vdash$) as a boolean relation, forgetting \emph{why} a sentence
holds. To persist witnesses alongside claims, Eigenius organizes hosted logics at
three levels~\cite{will2026theory}: (1)~at the \emph{witness level} (the commit
path), each reasoner is a Meseguer \emph{proof calculus}~\cite{meseguer1989}
assigning to each theory $T$ a category of witnesses $\mathcal{P}(T)$ projecting
onto sentences via $\pi_T$; (2)~forgetting witnesses yields a $\pi$-institution
at the \emph{consequence level}; and (3)~the canonical $\mathcal{G}$-construction
yields a standard institution satisfying the satisfaction condition at the
\emph{model level}.

Cross-system translations are \emph{maps of proof calculi} (comorphisms)
$(\Phi, \alpha, \gamma)$ executed inside the engine: $\Phi$ maps theories,
$\alpha_T$ translates sentences, and $\gamma_T : \mathcal{P}(T) \to
\mathcal{P}'(\Phi T)$ transports witnesses subject to the \emph{transport
invariant} $\pi'_{\Phi T}(\gamma_T(p)) = \alpha_T(\pi_T(p))$. Every comorphism
declares one of three \emph{axiom policies}~\cite{will2026theory}: \emph{Adopt}
(import source axioms), \emph{Discharge} (supply target proofs), or
\emph{Relativize} (prepend source axioms as hypotheses). Witness transport
composes pairwise along the DAG of registered comorphisms, whereas at the query
level EigenQL evaluates queries across all registered institutions simultaneously,
forming a common Grothendieck view~\cite{diaconescu2025} over the store's logics.

An institution registers itself via ontology resources, declaring its typed
\code{Export\-Format}s, \code{Import\-Format}s, and \code{Comorphism}s, and
interacts with the kernel through three trait methods: \code{extract\_typed},
\code{reify}, and \code{query}. A comorphism pipeline extracts a payload of
type $S$, applies $m : S \to T$, and reifies output $T$ into a target resource.
At commit, the kernel statically type-checks that $m$'s signature matches
\code{payload(export)}\,\allowbreak$\to$\,\code{payload(import)}; type-incorrect
translations are rejected like schema violations. Institutions run either
out-of-core in containers (Julia/R) or in-process inside the kernel binary (Lean~4).

\paragraph{Justification and statistics as pluggable logic.}
The epistemic grading from \S\ref{sec:datamodel} is not hardwired into the engine; 
justification logic itself is an attached institution. The replication in 
\S\ref{sec:impl} demonstrates this by composing two pluggable reasoners. First, 
a \emph{statistics institution} connects empirical computation to the commit path. 
It fires on a committed statistical-analysis plan---a recipe, not a claim:

\begin{quote}\small
\begin{verbatim}
resource wrn:wrn_dep_plan
       : stats:StatisticalAnalysisPlan {
  stats:sample_set     = wrn:wrn_dep_sampleset;
  stats:alpha          = 0.05;
  stats:directionality = OneSidedWitnessed(
      "urn:...:msi_directionality_witness");
  stats:variance_assumption = RankBased();
  stats:outlier_exclusion   = Identity();
}
\end{verbatim}
\end{quote}

\noindent No $p$-value and no conclusion appear anywhere in it;
even the choice of a one-sided test cites a chain resource rather
than being a free parameter. On commit, a verification trigger recomputes from
pinned data and emits two coupled \emph{derived} residents: a
verdict, and a result whose \code{canonical\_proposition} is
\code{stats:lt(stats:mean\_diff\_of(s),\,0)} --- the proposition
tested --- carrying an \code{IsDerivedAs} witness. The chain
commits a rerunnable procedure, not a bare claim. 

Second, a \emph{reasoning institution} discharges the
conclusions standing on those empirical results. Each conclusion is a 
proposition paired with a \emph{justification term} --- a typed accounting, 
in the sense of Artemov's justification logic~\cite{artemov-jl}, of how it was
reached. While the kernel provides the generic typing rule to discharge leaves 
(synthesizing the witness inhabitant from the layer's witness index, so a leaf 
naming no chain-resident warrant fails to type-check), the actual justification 
logic is just another institutional plugin. A \emph{derived} leaf is a statistics 
result, a \emph{declared} leaf a named bridge or rule; the four-warrant
taxonomy is thus a justification logic enforced by composition across the 
engine's query and commit paths.

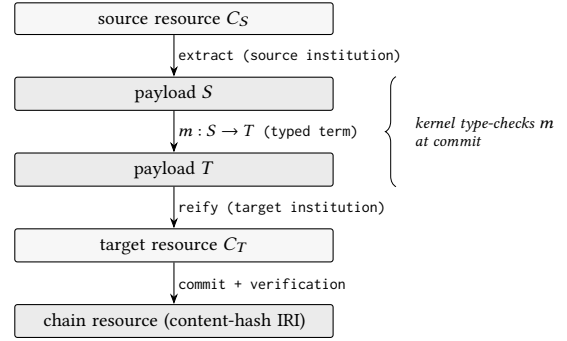
\begin{figure}[t]
\centering
\begin{tikzpicture}[
    font=\footnotesize,
    node distance=5.5mm,
    box/.style={draw, rounded corners=1pt, align=center,
      inner sep=3pt, fill=gray!6, line width=0.4pt, text width=40mm},
    op/.style={draw, rounded corners=1pt, align=center,
      inner sep=3pt, fill=gray!16, line width=0.4pt, text width=40mm},
    e/.style={-{Stealth[length=4pt]}, line width=0.4pt},
    lbl/.style={font=\scriptsize\ttfamily, inner sep=1.5pt}]
  \node[box] (cs) {source resource $C_S$};
  \node[op, below=of cs] (s) {payload $S$};
  \node[op, below=of s] (t) {payload $T$};
  \node[box, below=of t] (ct) {target resource $C_T$};
  \node[op, below=of ct] (chain)
    {chain resource (content-hash IRI)};
  \draw[e] (cs) -- node[lbl,right] {extract (source institution)} (s);
  \draw[e] (s) -- node[lbl,right] {$m : S \to T$ (typed term)} (t);
  \draw[e] (t) -- node[lbl,right] {reify (target institution)} (ct);
  \draw[e] (ct) -- node[lbl,right] {commit + verification} (chain);
  \draw[decorate,decoration={brace,amplitude=4pt},
        line width=0.3pt]
    ($(t.south east)+(8.25mm,0)$) -- ($(s.north east)+(8.25mm,0)$)
    node[midway,right=4pt,font=\scriptsize\itshape,text width=20mm,
         align=left] {kernel type-checks $m$ at commit};
\end{tikzpicture}
\caption{The comorphism pipeline. The kernel type-checks the
transformation $m$ against the export and import payload types at
commit; the reified target resource is then reinserted into the
chain as a first-class, content-addressed resident.}
\Description{Top-to-bottom pipeline: source resource of class C_S,
extract to payload S, transform via m from S to T (the kernel
type-checks m at commit), reify to target resource of class C_T,
then commit to a chain resource at a content-hash IRI.}
\label{fig:comorphism}
\end{figure}

\paragraph{Materializing the boundary: durable translations.}
In traditional federated databases or reasoning orchestrators (like 
Hets~\cite{mossakowski2007hets}), cross-system translations are 
ephemeral: computed in volatile memory to bridge a gap, then discarded. 
Because Eigenius treats both schemas and proofs as typed graph resources, 
a crossing's warranty---its typed translation---is just data. The reify 
step's output (Figure~\ref{fig:comorphism}) is not a transport-only 
intermediate; it is \emph{materialized} back into the chain at a deterministic 
content-hash IRI (\code{urn:eigenius:\allowbreak comorphism-output:\allowbreak\textit{tail}:\textit{hex}}), 
or at a caller-named IRI under an \code{INTO} clause. 
Identical translations naturally deduplicate via content addressing; 
commit-time triggers fire on the newly materialized translation; and 
downstream queries pattern-match it exactly like native data. The seam is 
crossed once, and the translation becomes a durable, verifiable graph resident.

\paragraph{Eliminating adapter cost via shared graph IRs.}
Polystore systems rely on pairwise adapters to move data between subsystems, 
creating an $O(N^2)$ integration bottleneck. Eigenius flattens this cost by 
lifting a shared, strongly typed Intermediate Representation (IR) directly into 
the graph's schema. When institutions agree on a payload drawn from the kernel's 
own type theory, the adapter degenerates to pure identity.

For example, the mathematical payload language is \code{FormulaTerm}---a typed 
AST modeled natively in the graph. The operator catalog is a set of chain resources, 
rank-checked at commit. All five Julia institutions (Symbolics, IntervalArithmetic, 
Catalyst, OrdinaryDiffEq, and JuMP-HiGHS) consume \code{FormulaTerm} as their 
native input. Because they share this graph-native IR, all comorphisms between 
them are identity transformations.

\principle{The $O(N^2)$ polystore integration cost is eliminated by modeling 
  a shared, strongly typed intermediate representation natively in the graph.}

\paragraph{EigenQL: typed Datalog with institution dispatch.} The
query language is a typed stratified Datalog with aggregation,
\code{DEFINE}d recursion, and two clauses particular to the institutional setting. 
\code{FIBER} dispatches execution into an external institution mid-query. \code{INTO} 
acts as an inline materialization step: it pins the external engine's response back 
into the chain at a named IRI, lifting it through the commit pipeline. The query
below matches docking results, coerces each through the
\code{dock\_to\_assay} comorphism inline, dispatches the coerced
value into the assay institution's validation query, and pins the
typed verdict back into the chain:

\begin{quote}\small
\begin{verbatim}
USING "urn:eigenius:dock:DockingResult"
USING INSTITUTION "urn:...:assay" AS assay

MATCH DockingResult(?d) { compound: ?c }
FIBER assay:validate_prediction {
    candidate: dock_to_assay(?d)   // inline comorphism
} AS ?v INTO "urn:eigenius:run:val_42"

RETURN [Validation] { compound: ?c, verdict: ?v }
\end{verbatim}
\end{quote}

\noindent The \code{INTO} target becomes a chain-resident
\code{Validation} resource at \code{urn:eigenius:\allowbreak run:val\_42}. 
Neither clause fits a traditional TVF: \code{FIBER} invokes complex typed 
reasoners outside the RDBMS, and \code{INTO} safely triggers validated 
materialization mid-evaluation.

\section{Formal Verification as a Database Invariant}
\label{sec:programs}

While empirical science captures its warrants via justification logic 
and \code{DerivedResource} commitments (\S\ref{sec:datamodel}), pure 
mathematics requires absolute proof. To enforce the stricter \code{VerifiedResource} 
grade, the database must ingest, store, and statically re-check the proof itself 
on the commit path. Doing so requires solving three architectural challenges.

\paragraph{Challenge 1: Proofs as native graph data and the proof-relevance boundary.}
EigenTT~\cite{minitt,will2026theory} represents a verified claim as a dependent
pair $(P, t)$, where $P : \mathrm{Prop}$ is expressed natively in EigenTT and
ontology classes resolve as ground types by walking the layer chain. Why not
translate a Lean~4 proof directly into EigenTT's $\mathrm{Prop}$? Beyond term
size, there is a \emph{proof-relevance mismatch}~\cite{will2026theory}: Lean~4's
$\mathrm{Prop}$ has \emph{definitional} proof irrelevance, whereas EigenTT keeps
$\mathrm{Prop}$ intensional (with only \emph{axiomatic} proof irrelevance) so
distinct warrants remain inspectable. A naive comorphism into a definitionally
proof-irrelevant sort either erases witness distinctions or violates conversion
preservation. Eigenius therefore encapsulates the exported Lean~4 derivation as
an external witness certificate (\code{Lean\-Proof\-Payload}, typed as
$\mathsf{Checked}(a)$) paired with $P = \iota\,T$ via the export comorphism $\iota$.

\paragraph{Challenge 2: The in-process commit gate.} 
The second challenge is validating this opaque witness $t$ without crippling 
throughput. Lean~4 acts strictly as an off-chain authoring environment. 
Calling out to an external orchestrator at commit would incur massive IPC 
overhead. Instead, Eigenius brings verification \emph{in-process}. The kernel's 
type checker runs the exported proof bytes through a minimal Rust term checker, 
\code{nanoda\_lib}~\cite{nanoda, carneiro2024}, linked directly into the database 
binary. Sharing the binary shortens the audit path without merging trust surfaces: 
a panic in the term checker is confined to the verification fiber and cannot corrupt 
the core ontology. Successful reasoning traces persist as chain resources, 
doubling as a native memoization cache.

\paragraph{Challenge 3: The binding problem.}
How does the database guarantee that an abstract Lean~4 theorem applies to the 
specific data stored in the graph? The chain shapes carrying a claim include a 
\code{LeanPackageMirror} (the source-layer anchor and mirrored mappings), a 
\code{LeanProofPayload} (the verbatim bytes), and a \code{LeanProofTerm}. 

A commit-time query fires on every \code{LeanProofTerm} and performs a strict 
\emph{three-part check}: (1) proof validity via the in-process term checker; 
(2)~mirror correspondence, mapping each \code{EigeniusFFI.*} constant in the 
proposition back to a chain class IRI (ensuring it references the schema); 
and (3) anchor consistency, hashing the mirror's archive against the declared 
content hash. All three must pass to yield \code{Verdict::Holds}.

\begin{figure}[t]
\centering
\resizebox{\columnwidth}{!}{%
\begin{tikzpicture}[
    font=\footnotesize,
    node distance=5mm and 7mm,
    res/.style={draw, rounded corners=1pt, align=center,
      inner sep=2.5pt, fill=gray!6, line width=0.4pt},
    cls/.style={draw, rounded corners=1pt, align=center,
      inner sep=2.5pt, fill=gray!16, line width=0.4pt},
    e/.style={-{Stealth[length=4pt]}, line width=0.4pt},
    lbl/.style={font=\scriptsize\ttfamily, inner sep=1.2pt}]
  \node[res] (v) {Verdict(``Holds'')};
  \node[res, below=of v] (pt) {LeanProofTerm};
  \node[cls, right=18mm of pt] (pat) {patient\_1 : Patient};
  \node[res, below left=7mm and -6mm of pt] (pay)
    {LeanProof\-Payload\\\scriptsize(export bytes)};
  \node[res, below right=7mm and -6mm of pt] (mir) {LeanPackageMirror};
  \node[res, below=14mm of mir] (lay) {bootstrap head layer};
  \draw[e] (v) -- node[lbl,right] {verdict\_subject} (pt);
  \draw[e] (pt) -- node[lbl,above] {claim\_iri} (pat);
  \draw[e] (pt) -- node[lbl,left,pos=0.55] {proof\_payload} (pay);
  \draw[e] (pt) -- node[lbl,right,pos=0.55] {mirror\_iri} (mir);
  \draw[e] (mir) -- node[lbl,right] {source\_layer} (lay);
  \draw[e] (mir.east) to[out=25,in=-90]
    node[lbl,right,pos=0.7] {mirrored\_classes} (pat.south);
  \node[font=\scriptsize\itshape, text width=34mm, align=left,
        right=3mm of mir.east |- lay] (closes)
    {claim\_iri and mirrored\_classes converge on the same
     class: the cycle closes};
\end{tikzpicture}%
}
\caption{The closed audit chain: every edge is a
content-addressed chain reference, and the cycle from
\code{Verdict} back to the ontology class is re-walkable and
re-checkable from the graph alone.}
\Description{Directed graph of the audit cycle from Verdict
through LeanProofTerm, LeanProofPayload, and LeanPackageMirror
(with mirrored_classes, content hash, source layer) and via
claim_iri to the patient_1 instance and its Patient class,
closing the cycle.}
\label{fig:audit}
\end{figure}
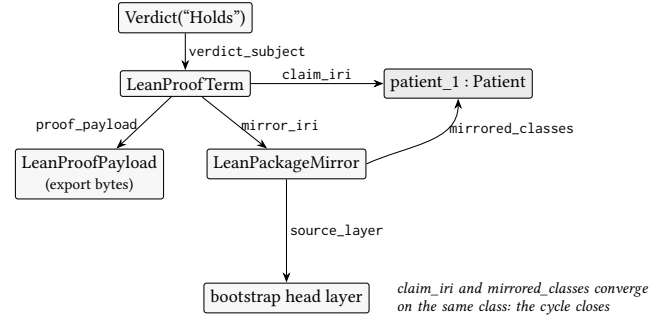

\paragraph{The closed audit chain.}
This architecture closes the audit loop within the database (Figure~\ref{fig:audit}). 
An auditor walks from a \code{Verdict} to the proof term, payload, and chain class, 
re-fetching every byte from the content-addressed store. They can re-run the 
in-process check with no contact to intermediate middleware. To mark a resource 
\emph{verified} is not to ask the user to trust the database software, but to 
store the machine-checkable mathematics as a queryable chain resident.

\section{Implementation and Evaluation}
\label{sec:impl}

\paragraph{Prototype and baseline flows.} The Rust prototype includes RocksDB 
storage, a gRPC API, and a React notebook. An MCP server~\cite{mcp} exposes 
kernel ops to autonomous agents. Eight institutions run live across two extensibility 
paths: five Julia institutions on the runtime substrate, Lean~4 in-process, and 
the statistics and reasoning institutions. 

To demonstrate that the kernel supports both domains of scientific epistemology, 
the repository contains two baseline flows. The mathematical path is exercised 
by a Lean-verification flow whose EigenQL queries walk the closed audit chain from 
a \code{Verdict::Holds} back to the chain class. The empirical path is demonstrated 
by a kinase-screening flow that lifts 24 IC$_{50}$ measurements to typed reasoning 
through identity comorphisms across the Julia stack. 

\paragraph{Microbenchmark: Typed ingestion at scale.} 
To evaluate the overhead of treating every record as a typed, commit-verified 
resource, we ingest the WordNet and UMLS lexicons. This data materializes as 
9.19M chain resources at 361\,bytes each (6.6M lexical entries, 4.8M surface forms). 
Import takes the standard commit path with no bulk-load bypass, sustaining roughly 
5{,}200 validated resources/s on one core of a 22-CPU laptop. The commit gate is 
where the cost lies: structural class layers load at ${\sim}28$k resources/s, while 
lexical entries, which rigorously type-check each sense against its category, load 
at ${\sim}4$k resources/s.

\paragraph{Macro-evaluation: Replicating a published study.} The
substrate's flagship evaluation scales the empirical path to an end-to-end 
recomputation of a high-profile \emph{Nature} study --- Chan et al.~\cite{chan2019} 
(WRN helicase in microsatellite-unstable cancers). We chose it as a stress test 
because its argument crosses multiple domains: genome-wide dependency screens, 
wet-lab validation, in vivo models, and a DNA-damage mechanism.

From content-addressed, hash-pinned source data, two institutions compose over 
one layer chain. A \emph{statistics institution} re-runs each statistic and 
commits a derived result with an \code{IsDerivedAs} witness. A \emph{reasoning 
institution} reads those witnesses and discharges the study's domain conclusions, 
each a \code{ReasoningSentence} whose certificate is a justification term 
type-checked against the asserted proposition. Every claim carries one of four 
warrant grades, making its epistemic status a queryable property of the chain 
rather than a prose caveat. On a clean database, \textbf{52 of 52 derived 
conclusions hold}, discharging the study's conclusions from pinned data. 

The value of this massive structural re-encoding is what it surfaces: 
recomputation exposed four discrepancies between the paper's prose and its own 
data, including a corrected sample size ($n = 54 \to 51$, rows dropped as missing) 
and a pseudoreplication whose honest $p$-value differs by thirteen orders of 
magnitude. These are recorded not as footnotes but as machine-checkable facts 
carrying both warrants side by side. Forcing every bridge to be a \emph{declared} 
resource makes the surface of assumptions inspectable by an agent or human. 
Every analysis class in the paper runs live---from a \code{limma} differential-dependency 
call over a 187\,MB matrix to per-cell mechanism assays---and the few readouts 
that remain merely cited say so on the chain.

\paragraph{Next step: the narrative itself.} The replication formalizes the study's 
steps and deviations; the prose argument connecting them stays off the chain. We are 
closing that gap with a dependently-typed categorial grammar engine that lands the 
narrative as typed chain-resident terms: prose is paraphrased into controlled English 
--- one statement per sentence --- then parsed compositionally, following 
MTT-semantics~\cite{mtt-semantics} and the CCG-with-dependent-types lineage of 
lightblue~\cite{lightblue}. The controlled-English stage provides a narrow target 
the model produces reliably and a categorial grammar parses deterministically. 
Because the terms land in the kernel's own type theory, the argument --- not just 
its steps --- becomes a chain resident under the same institution checks as the 
data it describes. The engine is headed for registration as a further institution, 
and because full EigenTT serves as a cross-domain representation, the identity 
collapse above extends to prose: one more logic in the diagram, not a layer above it.

\section{Discussion}
\label{sec:discussion}

\paragraph{Related work.} 
Eigenius synthesizes ideas across database and formal-methods traditions by pulling 
their mechanisms into the kernel. 
\emph{Versioning and federation:} Versioned stores like Dolt~\cite{dolt} 
and TerminusDB~\cite{terminusdb} share immutable history, but reconcile untyped 
rows. Eigenius's merges are strongly typed functorial migrations~\cite{spivak2012}. 
Polystore systems integrate heterogeneous storage, but rely on adapters rather 
than native comorphisms that reinsert results into the chain. 
\emph{Provenance and shapes:} Provenance vocabularies~\cite{provdm} and 
graph shapes~\cite{shacl} (over RDF~\cite{rdf} and Atomic Data~\cite{atomicdata}) 
model audit relationships \emph{post hoc}; Eigenius enforces stratification as a 
commit-time invariant. 
\emph{AI and orchestration:} Hets~\cite{mossakowski2007hets} orchestrates institutions 
ephemerally. Finally, neurosymbolic lifting~\cite{pareschi2026} reconstructs LLM
outputs compositionally and treats the trustworthiness of a knowledge source as an
open meta-question, isolated behind a contract-and-oracle interface; Eigenius resolves
that question by absorbing the oracle into the auditable substrate, making source
trust a re-walkable property of the chain rather than an opaque assumption.

\paragraph{What we got wrong.} The institution protocol was originally built
around Wasm components, assuming strict sandboxing was the right boundary. 
While theoretically elegant, real-world reasoners (ODE solvers, biostatistics) 
wrap heavy native libraries that do not compile cleanly to Wasm. We abandoned 
Wasm for sibling containers because supporting scientific ecosystems proved more 
important than lightweight sandboxing. Additionally, the kernel/orchestrator 
boundary remains in flux and represents ongoing technical debt.

\paragraph{Mechanized metatheory and open problems.}
The core theoretical foundations of Eigenius are established and mechanized in
Lean~4 in our companion monograph~\cite{will2026theory} (with zero unproved
obligations): confluence, injectivity, and separation of conversion ($\equiv$)
from ontology subsumption ($\leq$); soundness, completeness, and set-theoretic
consistency of the bidirectional kernel $\mathcal{K}_0/\mathcal{K}_1$ and typed
knowledge-graph calculus; the justification logic $\mathcal{J}$ and its
four-grade lattice; the three-level institutional framework and its axiom
policies; and soundness and completeness of untyped NbE and the algorithmic
checker $\mathcal{A}_0/\mathcal{A}_1$.

Three foundational and systems questions remain open. First, termination of
$\mathcal{A}_0/\mathcal{A}_1$ is proved relative to a strong-normalization
hypothesis for well-typed terms; discharging strong normalization
unconditionally for the full kernel theory and verifying the production Rust
kernel against the Lean~4 specification remain open. Second, while additive
extensions, renamings, and conservative projections preserve typing
functorially, characterizing general pushouts in $\mathbf{Th}$ across
conflicting schema-and-instance branches remains open. Finally, beyond the
\emph{Nature} replication, the decisive systems test is closing the loop:
proving that an AI Scientist operating strictly over MCP endpoints can
autonomously traverse, mutate, and verify a scientific graph without human intervention.

\begin{acks}
This work builds on the ideas behind the original Pharos research artifact, created
at Microsoft Research's Health Solutions Group; the authors thank
Rick Sax, Andy McGregor, and the wider Pharos team for that foundation.
\end{acks}

\bibliographystyle{plainnat}

\begingroup\footnotesize
\begin{thebibliography}{30}

\bibitem[Brown(2008)]{brown2008pharos}
A.L.~Brown, Jr.
\newblock Enforcing the scientific method.
\newblock In \emph{Provenance and Annotation of Data and
  Processes}, page~2. Springer, 2008.

\bibitem[Microsoft(2009)]{amalga}
Microsoft.
\newblock Microsoft introduces groundbreaking technology for life
  sciences.
\newblock Press release for {Amalga Life Sciences 2009}, Apr.~28,
  2009.
\newblock
  \url{https://news.microsoft.com/source/2009/04/28/microsoft-introduces-groundbreaking-technology-for-life-sciences/}.

\bibitem[Will(2026)]{will2026theory}
H.-M.~Will.
\newblock \emph{Typed Knowledge over Constructive Institutions: {Eigenius},
  Volume~I: Theory}.
\newblock The Eigenius Project, 2026.
\newblock \url{https://doi.org/10.5281/zenodo.23964990}.

\bibitem[Goguen and Burstall(1992)]{goguen1992}
J.A.~Goguen and R.M.~Burstall.
\newblock Institutions: Abstract model theory for specification
  and programming.
\newblock \emph{JACM}, 39(1):95--146, 1992.

\bibitem[Meseguer(1989)]{meseguer1989}
J.~Meseguer.
\newblock General logics.
\newblock In \emph{Logic Colloquium~'87}, Studies in Logic and the
  Foundations of Mathematics 129, pp.~275--329. Elsevier, 1989.

\bibitem[Fiadeiro and Sernadas(1988)]{fiadeiro-sernadas1988}
J.~Fiadeiro and A.~Sernadas.
\newblock Structuring theories on consequence.
\newblock In \emph{Recent Trends in Data Type Specification}, LNCS 332,
  pp.~44--72. Springer, 1988.

\bibitem[Diaconescu(2025)]{diaconescu2025}
R.~Diaconescu.
\newblock \emph{Institution-independent Model Theory}.
\newblock Springer, 2nd edition, 2025.

\bibitem[Spivak(2012)]{spivak2012}
D.I.~Spivak.
\newblock Functorial data migration.
\newblock \emph{Information and Computation}, 217:31--51, 2012.

\bibitem[Bormann and Hoffman(2020)]{rfc8949}
C.~Bormann and P.~Hoffman.
\newblock Concise Binary Object Representation ({CBOR}).
\newblock RFC 8949, Request for Comments, IETF, 2020.

\bibitem[Mossakowski et~al.(2007)]{mossakowski2007hets}
T.~Mossakowski, C.~Maeder, and K.~L\"{u}ttich.
\newblock The heterogeneous tool set, {Hets}.
\newblock In \emph{TACAS 2007}, LNCS 4424, pp.~519--522. Springer,
  2007.

\bibitem[Coquand et~al.(2009)]{minitt}
T.~Coquand, Y.~Kinoshita, B.~Nordstr\"{o}m, and M.~Takeyama.
\newblock A simple type-theoretic language: {Mini-TT}.
\newblock In \emph{From Semantics to Computer Science},
  pp.~139--164. CUP, 2009.

\bibitem[Abel et~al.(2007)]{nbe}
A.~Abel, T.~Coquand, and P.~Dybjer.
\newblock Normalization by evaluation for {Martin-L\"of} type
  theory with typed equality judgements.
\newblock In \emph{LICS 2007}, pp.~3--12. IEEE, 2007.

\bibitem[de~Moura and Ullrich(2021)]{lean4}
L.~de~Moura and S.~Ullrich.
\newblock The {Lean~4} theorem prover and programming language.
\newblock In \emph{CADE-28}, LNCS 12699, pp.~625--635. Springer,
  2021.

\bibitem[Carneiro(2024)]{carneiro2024}
M.~Carneiro.
\newblock {Lean4Lean}: Verifying a typechecker for {Lean}, in
  {Lean}.
\newblock arXiv:2403.14064, 2024.

\bibitem[Bailey(2024)]{nanoda}
C.~Bailey.
\newblock {nanoda\_lib}: A {Lean~4} term checker library in {Rust}.
\newblock \url{https://github.com/ammkrn/nanoda_lib}.

\bibitem[Cyganiak et~al.(2014)]{rdf}
R.~Cyganiak, D.~Wood, and M.~Lanthaler.
\newblock {RDF 1.1} concepts and abstract syntax.
\newblock W3C Recommendation, 2014.

\bibitem[Meindertsma(2020)]{atomicdata}
J.~Meindertsma.
\newblock {Atomic Data}.
\newblock W3C Community Group Specification, 2020--2026.

\bibitem[Sanca and Heyse(2022)]{dolt}
V.~Sanca and T.~Heyse.
\newblock {Dolt}: {SQL} with {Git}-like versioning.
\newblock DoltHub, 2019--2026.

\bibitem[Artemov(2008)]{artemov-jl}
S.~Artemov.
\newblock The logic of justification.
\newblock \emph{The Review of Symbolic Logic}, 1(4):477--513, 2008.

\bibitem[TerminusDB Team(2024)]{terminusdb}
TerminusDB Team.
\newblock {TerminusDB}: A delta-encoded graph database with
  revision control.
\newblock \url{https://terminusdb.com/}, 2024.

\bibitem[Moreau et~al.(2013)]{provdm}
L.~Moreau et~al.
\newblock {PROV-DM}: The {PROV} data model.
\newblock W3C Recommendation, 2013.

\bibitem[Knublauch and Kontokostas(2017)]{shacl}
H.~Knublauch and D.~Kontokostas.
\newblock Shapes constraint language ({SHACL}).
\newblock W3C Recommendation, 2017.

\bibitem[Anthropic(2024)]{mcp}
Anthropic.
\newblock {Model Context Protocol}.
\newblock Open specification, 2024--2026,
  \mbox{\url{https://modelcontextprotocol.io}}.

\bibitem[Chan et~al.(2019)]{chan2019}
E.M.~Chan, T.~Shibue, J.M.~McFarland, et~al.
\newblock {WRN} helicase is a synthetic lethal target in
  microsatellite unstable cancers.
\newblock \emph{Nature}, 568:551--556, 2019.
\newblock \url{https://doi.org/10.1038/s41586-019-1102-x}.

\bibitem[Pareschi(2026)]{pareschi2026}
R.~Pareschi.
\newblock From dependency to compositionality: A neurosymbolic
  lifting of {LLM} outputs via combinatory categorial grammar.
\newblock arXiv:2607.18961, 2026.

\bibitem[Brown(2026)]{brown2026pct}
A.L.~Brown, Jr.
\newblock \emph{Distributed decentralized clinical trial: Design to
  executable digital twin}.
\newblock PCT International Patent Application
  No.\ \texttt{PCT/US2026/018207}, filed Mar.~6, 2026. Patent
  pending; international publication forthcoming.

\bibitem[Chatzikyriakidis and Luo(2020)]{mtt-semantics}
S.~Chatzikyriakidis and Z.~Luo.
\newblock \emph{Formal Semantics in Modern Type Theories}.
\newblock Wiley-ISTE, 2020.

\bibitem[Tomita et~al.(2025)]{lightblue}
A.~Tomita et~al.
\newblock Natural language inference with a {CCG} parser and
  automated theorem prover for {DTS} ({lightblue}/{wani}).
\newblock In \emph{BriGap-2}, pp.~1--7, 2025.

\end{thebibliography}
\endgroup
\renewcommand{\bibfont}{\scriptsize}
\setlength{\bibsep}{0pt plus 0.3pt}

\end{document}